\documentclass[runningheads,a4paper]{llncs}

\usepackage{amssymb}
\usepackage{graphicx}

\usepackage[latin1]{inputenc}
\usepackage{tikz}
\usetikzlibrary{shapes,arrows, positioning}
\pgfmathtruncatemacro\distance{1}

\usepackage{url}
\urldef{\mailsa}\path|amironov66@gmail.com|
\urldef{\mailsb}\path|anna.kramer, leonie.kunz, christine.reiss, nicole.sator,|
\urldef{\mailsc}\path|erika.siebert-cole, peter.strasser, lncs}@springer.com|    
\newcommand{\keywords}[1]{\par\addvspace\baselineskip
\noindent\keywordname\enspace\ignorespaces#1}

\begin{document}

\def\simplus{
\approx
}

\def\ite#1#2#3{\left\{\begin{array}{rlll}
[\![{#1}]\!]&{#2}\\
:&{#3}
\ey\right.}

\def\g#1{\left[\!\!\left[
{\def\arraystretch{1}
\begin{array}{lllll}#1\end{array}}
\right]\!\!\right]}

\def\bcy{\begin{array}{ccc}}

\def\dey#1#2{#1 (#2)}
\def\deyc#1#2{#1 \cdot  #2}
\def\bcy{\begin{array}{ccc}}

\def\ral#1{\;\mathop{\longrightarrow}\limits^{#1}\;}
\def\bc{\begin{center}\begin{tabular}{l}}
\def\ec{\end{tabular}\end{center}}
\def\modn#1{\mathop{=}\limits_{#1}}
\def\modnop#1{\mathop{#1}\limits_{n}}
\def\rar{\mathop{\in}\limits_{r}}
\def\und#1{\mathop{=}\limits_{#1}}
\def\skobq#1{\langle\!| #1 |\!\rangle}
\def\redeq{\;\mathop{\approx}\limits^{r}\;}
\def\reduc{\;\mathop{\mapsto}\limits^{r}\;}
\def\pt{\;\mathop{+}\limits_{\tau}\;}
\def\sost{\begin{picture}(0,0)\put(0,3){\circle*{4}}
\end{picture}}
\def\sosto{\begin{picture}(0,0)\put(0,0){\circle*{4}}
\end{picture}}
\def\bi{\begin{itemize}}
\def\pa{\,|\,}
\def\oc{\;\mathop{\approx}\limits^{+}\;}
\def\p#1#2{(\;#1\;,\;#2\;)}
\def\mor#1#2#3{\by #1&\pright{#2}&#3\ey}
\def\ei{\end{itemize}}
\def\bn{\begin{enumerate}}
\def\en{\end{enumerate}}
\def\i{\item}
\def\a{\forall\;}

\def\l#1{[#1]}
\def\lc#1{\langle#1\rangle}
\def\lcc#1#2{{#1}^{#2}}

\def\fm#1{\left[\!\!\left[\begin{array}{lllll}#1\end{array}\right]\!\!\right]}
\def\ra#1{\mathop{\to}\limits^{\!\!#1}}
\def\dra#1{\mathop{\Rightarrow}\limits^{\!\!\!\!#1}}

\def\bigset#1#2{\left\{\by #1 \left| \by #2 \ey\right\}\ey\right.}
\def\p{\leftarrow}

\def\plongright#1{
  \begin{picture}(40,18)
  \put (-5,3){\vector(1,0){50}}
  \put (20,8){\makebox(1,1){$\scriptstyle #1$}}
  \end{picture} }

\def\plongleft#1{
  \begin{picture}(40,8)
  \put (45,3){\vector(-1,0){50}}
  \put (20,8){\makebox(1,1){$\scriptstyle #1$}}
  \end{picture} }

\def\pse#1#2{
  \begin{picture}(40,8)
  \put (-5,-5){\vector(2,-1){50}}
  \put (45,-5){\vector(-2,-1){50}}
  \put (-5,-12){\makebox(1,1)[r]{$\scriptstyle #1$}}
  \put (45,-12){\makebox(1,1)[l]{$\scriptstyle #2$}}
  \end{picture} }

\def\und#1{\mathop{=}\limits_{#1}}
\def\redeq{\;\mathop{\approx}\limits^{r}\;}
\def\reduc{\;\mathop{\mapsto}\limits^{r}\;}
\def\oc{\mathop{\approx}\limits^{+}}
\def\sost{\begin{picture}(0,0)\put(0,0){\circle*{4}}
\end{picture}}
\def\bi{\begin{itemize}}
\def\pa{\,|\,}
\def\oo{\;\mathop{\approx}\limits^{c}\;}
\def\p#1#2{(\;#1\;,\;#2\;)}
\def\mor#1#2#3{\by #1&\pright{#2}&#3\ey}
\def\ei{\end{itemize}}
\def\bn{\begin{enumerate}}
\def\en{\end{enumerate}}
\def\i{\item}
\def\bigset#1#2{\left\{\by #1 \left| \by #2 \ey\right\}\ey\right.}
\def\p{\leftarrow}
\def\buffer{{\it Buffer}}
\def\eam{\mathbin{{\mathop{=}\limits^{\mbox{\scriptsize def}}}}}
\def\be#1{\begin{equation}\label{#1}}
\def\ee{\end{equation}}
\def\re#1{(\ref{#1})}

\def\bn{\begin{enumerate}}
\def\en{\end{enumerate}}
\def\bi{\begin{itemize}}
\def\ei{\end{itemize}}
\def\i{\item}
\def\c#1{\left\{\begin{array}{lllll}#1\end{array}\right\}}
\def\d#1{\left[\begin{array}{lllll}#1\end{array}\right]}
\def\b#1{\left(\begin{array}{lllll}#1\end{array}\right)}
\def\ra#1{\;\mathop{\to}\limits^{\!\!#1}\;}
\def\leqd{\;\mathop{<}\limits_{2}\;}
\def\diagrw#1{{
  \def\normalbaselines{\baselineskip20pt \lineskip3pt \lineskiplimit3pt }
  \matrix{#1}}}

\def\blackbox{\vrule height 7pt width 7pt depth 0pt}
\def\pu#1#2{
\mbox{$\!\!\begin{picture}(0,0)
\put (-#1,-#2){\line(1,0){#1}}
\put (-#1,-#2){\line(0,1){#2}}
\put (#1,#2){\line(-1,0){#1}}
\put (#1,#2){\line(0,-1){#2}}
\put (-#1,#2){\line(1,0){#1}}
\put (-#1,#2){\line(0,-1){#2}}
\put (#1,-#2){\line(-1,0){#1}}
\put (#1,-#2){\line(0,1){#2}}
\end{picture}$}
}

\def\pright#1{
  \begin{picture}(20,18)
  \put (-5,3){\vector(1,0){30}}
  \put (9,8){\makebox(1,1){$\scriptstyle #1$}}
  \end{picture} }

\def\by{\begin{array}{llllllllllllll}}
\def\ey{\end{array}}

\mainmatter  

\title{A new method of verification \\of security protocols}

\author{Andrew M. Mironov
}

\institute{Moscow State University\\
Faculty of Mechanics and Mathematics
\\$\;$\\
\mailsa\\
}

\maketitle

\begin{abstract}
In the paper we introduce a process model of security protocols,
where processes are graphs
with edges labelled by 
  actions,
 and present a new method of specification and verification of security protocols based on this model.
\end{abstract}


\section{Introduction}

\subsection{Security protocols 
and their properties}

A {\bf security protocol}
({\bf SP}) is a 
distributed algorithm 
that determines an order 
of message passing
between several agents.
Examples of such agents 
are computer systems, 
bank cards, people, etc.
Messages transmitted 
by SPs can be encrypted.
We  assume
that encryption 
transformations 
used in SPs 
are perfect, i.e. satisfy
some axioms 
expressing, for example,
an impossibility of 
extraction of open texts from 
ciphertexts 
without knowing the corresponding cryptographic keys.

In this paper we present 
a new model of SPs
based on Milner's 
Calculus of Communicating
Systems \cite{milner}
and theory of processes
with message passing
\cite{tproc}. This model is 
a graph analog of a Calculus 
of Cryptographic Protocols 
({\bf spi-calculus}, 
\cite{82}). 
It can serve as a theoretical 
foundation for 
a new method
(presented in the paper)
of {\bf verification} of SPs, 
where verification means
a constructing of 
mathematical proofs that
SPs meet the desired properties.
Examples of such properties 
are {\bf integrity} 
and {\bf secrecy}.
These properties are defined formally, as some conditions
expressed in terms of
an observational
equivalence.

\subsection{Verification 
of security protocols}

There are examples of SPs
(\cite{1}--\cite{15})
which were used
in safety-critical systems, 
however it turned out that 
the SPs contain vulnerabilities 
of the following forms:
\bi
\i  agents involved in these
SPs
can receive
distorted messages (or lose them)
as a result of interception,
deletion or distortion of 
transmitted messages by an adversary, 
that violates 
the integrity property,
\i an adversary  can find out
a confidential information 
contained
in intercepted messages
as a result of erroneous
or fraudulent
actions of SP agents.
\ei

These examples justify that
for SPs used in 
safety-critical systems
it is not enough informal 
analysis of required properties, 
it is necessary 
\bi
\i to build  a 
{\bf mathematical model}
  of an analyzed SP,
\i to describe security 
  properties 
  of the analyzed SP
  as mathematical objects (e.g.
graphs, or logical formulas),
  called a {\bf formal specification}, and 
\i to construct a mathematical
proof that the analyzed SP
meets (or does not meet) 
the formal specification,
this proof is
called a {\bf formal verification}.
\ei

In the process model
described in the paper
SPs and 
their formal specifications 
are represented by 
processes with 
message passing. 
Many important properties 
of SPs (in particular, 
integrity and secrecy) 
can be expressed as 
observational equivalence 
of such processes.

One of the most significant advantages of the proposed
process model of SPs is 
a low complexity of proofs
of correctness of SPs.
In particular, there is 
no need to build a set of 
all reachable states 
of analyzed SPs, 
if the set of 
all these states 
and transmitted messages
is unbounded.

Among other  models of SPs
most popular  are logical models
(\cite{4}--
\cite{27}). These models 
provide possibility
to reduce the problem of 
verification of SPs
 to the problem
of proofs of theorems that 
analyzed SPs meet their 
specifications.
Algebraic and logical approaches to verification are considered also in 
 \cite{rsm}--\cite{corin}.

\section{Description of a process model of security protocols}

In the process model 
described below
SPs and formal specifications
of their properties 
are represented by graphs, 
whose edges
are labeled by  {\bf actions}.
Actions are expressions 
consisting of terms and formulas.

\subsection{Variables, constants,
terms}

We assume that there are given
a set
${\cal X}$ of {\bf variables}, 
 a subset ${\cal K}\subseteq 
{\cal X}$ of 
{\bf keys}, and 
a set
${\cal C}$ 
of {\bf constants}.
A set  ${\cal E}$ of {\bf terms}
is defined inductively:
\bi
\i 
$\forall\,x\in {\cal X},\;\forall\,c\in {\cal C}\quad
x$ and $c$ are terms,
\i for each list
$e_1,\ldots, e_n$ 
of 
terms the record  
$e_1\ldots e_n$
is a term, \\
(if the above list is empty, then the corresponding term is denoted by $\varepsilon$), 
\i $\forall\,k\in {\cal K}$, 
$\forall\, e\in {\cal E}$ the record 
$k(e)$ is a term
(called an {\bf encrypted message (EM)}, this term
represents
a result of an encryption
of $e$
on the key $k$).
\ei

Terms are designed for a representation of  messages transmitted between participants of communications,
a term of the form $e_1\ldots e_n$ represents 
a composite message consisting of  
messages corresponding to the 
components
$e_1,\ldots, e_n$.
$\forall\,e\in {\cal E}$ 
the set  of variables
occurred in $e$ is 
denoted by 
$X_e$. 
If terms $e,e'$ have the form $e_1,\ldots, e_n$
and $e'_1,\ldots, e'_{n'}$, respectively, then 
  the record $ee'$ denotes the term
$e_1,\ldots, e_n
e'_1,\ldots, e'_{n'}$, and $\forall\,e\in {\cal E}\;\;\varepsilon e =
e\varepsilon=e$.

%
%
%
%

\subsection{Formulas}

{\bf Elementary formulas (EFs)}
are records of the form 
$e=e'$ and
$e\in E$  
(where
$e,e'\in {\cal E}$,
and $E$ is a  subset of 
${\cal E}$).
A {\bf formula} is a conjunction of EFs.
The symbols $\top$ and $\bot$
denote true and false formulas respectively
(for example, $\top=(c_1=c_1)$, 
$\bot=
(c_1=c_2)$,
where $c_1$ and $c_2$ 
are different constants).
A set of formulas is denoted by
${\cal B}$.
$\forall\,b\in {\cal B}\;\;
X_b$ is a  set of all  variables 
occurring in $b$. 

$\forall\,b_1,b_2\in {\cal B}\quad
b_1\leq b_2$
means that $b_2$
is a logical   consequence of 
$b_1$ (where 
the concept of a 
logical consequence is defined by a 
standard way).

If $b_1\leq b_2$ and $b_2\leq b_1$, 
then 
 $b_1$ and $b_2$
are assumed to be equal.


$\forall\,k,k'\in {\cal K},\;\forall\,e,e'\in {\cal E}$
the formulas 
$k(e)=k'(e')$ and 
$(k=k')\wedge (e=e')$ are assumed to be 
equal.
The records $e_1=_be_2$
and $b\in_b E$
means that
$b\leq (e_1=e_2)$
and $b\leq (e\in E)$
respectively.

\subsection{Closed sets of terms}

Let $E\subseteq {\cal E}$ and $b\in {\cal B}$.
The set $E$  is said to be 
 $b$--{\bf closed} 
if 
\bi\i $\big(\forall\,i=1,\ldots,n\;\;
e_i\in E\big)\Leftrightarrow
e_1\ldots e_n\in E$,
\i $\forall\, k\in E\cap{\cal K}\;\;
\big(e\in E\Leftrightarrow
k(e)\in E\big)$,
\i $\forall\,e,e'\in {\cal E}\quad
(e=_be')\;\Rightarrow\;
\big(e\in E \;\Leftrightarrow\;e'\in E\big).$
\ei
Closed sets of terms are used for representation of sets of messages which can be known to an adversary. 
The above conditions 
correspond to operations 
which an adversary $A$
can perform 
 with his available messages:
\bi
\i if $A$ has  $e_1,\ldots, e_n$, then
it can compose the message
$e_1\ldots e_n$,
\i if $A$ has 
$e_1\ldots e_n$, 
then it may
get its components
$e_1$,
$\ldots$, $e_n$,
\i if $A$ has  
$k$ and $e$, 
where $k$ is a key, 
then it can create a EM 
$k(e)$, 
\i if $A$ has  
an EM $k(e)$ and
a key
$k$,  then
it can decrypt $k(e)$, i.e. 
get  $e$.
\ei

\refstepcounter{theorem}
{\bf Theorem \arabic{theorem}\label{threclo}.}
$\forall\,E\subseteq {\cal E}$, $\forall\,b\in {\cal B}$
there is a least (w.r.t. an inclusion of sets) $b$--closed set
$E^b\subseteq {\cal E}$,
such that  $E\subseteq E^b$. $\blackbox$\\

Let $D_1,D_2\subseteq
{\cal E}$, and 
$b_1,b_2\in {\cal B}$.
A binary relation
$\mu \subseteq 
D_1^{b_1}\times
D_2^{b_2}$
is said to be a
{\bf similarity}
between 
$(D_1,{b_1})$ and
$(D_2,{b_2})$, if 
$\forall\,(e_1,e_2)\in \mu$
\bi
\i $\forall\,
e'_1,e'_2\in {\cal E}\quad
(e'_1,e_2)\in \mu\Leftrightarrow
(e_1=_{b_1}e'_1),\;\;
(e_1,e'_2)\in \mu\Leftrightarrow
(e_2=_{b_2}e'_2)$,
\i the conditions
$\exists\;
e_{i}^1,\ldots ,e_{i}^n \in D_i^{b_i}: 
(e_i=_{b_i}e_{i}^1\ldots e_{i}^n)\;\;
(i=1,2)$ are equivalent,
and if these 
conditions hold, then
$\forall\,i=1,\ldots, n\;\;
(e_{1}^i,e_{2}^i)\in\mu$,
\i the conditions 
$\exists \,k_{i},e_{i}'\in D_i^{b_i}:
 (e_i=_{b_i}k_{i}(e_{i}'))\;\;
(i=1,2)$ are equivalent,
and if these 
conditions hold, then
$(k_{1},k_{2})\in\mu$ and $(e_{1}',e_{2}')\in\mu$.
\ei
 
A set of all 
similarities
between 
$(D_1,{b_1})$ and
$(D_2,{b_2})$
is denoted by the record
$Sim\big((D_1,{b_1}),
(D_2,{b_2})\big)$.


\subsection{Actions}
An {\bf action} is a record 
of one 
of the three 
kinds: an input, an output, an internal action.
Inputs and outputs 
are associated with 
an   
{\bf execution}, 
defined below.
\bi
\i
An {\bf input} is an action of the form
$e?e'$, where 
$e,e'\in {\cal E}$.
An execution 
of this action 
consists of a receiving
a message through a channel
named 
$e$, and 
writing components of this
 message
to  variables occurring in 
$e'$.

\i An {\bf output} is an action of the form $e! e'$, where $e,e'\in  {\cal E}$.
An execution of this action
consists of a sending a message $e'$
through a channel named   
$e$.

\i An {\bf internal action}
is an action of the form
$b$, where $b\in {\cal B}$.
\ei

The set of all actions
is denoted by
${\cal A}$,   
$\forall\,a\in {\cal A}$ 
a set of  variables
occurred in $a$
is denoted by 
$X_{a}$.

\subsection{Processes
with a message passing}

Processes with a message passing
are intended for description
of SPs and formal
specifications of their 
properties.

A {\bf process
with a message passing} 
(called below briefly as a {\bf process}) is a tuple
$P=(S, s^0,R,b^0,
 D^0, H^0)$, where
\bi\i $S$ is a set of {\bf states},
$s^0 \in S$ is an {\bf initial state},
\i $R\subseteq S\times {\cal A}\times S$ is a set of {\bf transitions},  
each transition
$(s,a,s')\in R$
is denoted by
the record $s\ra{a}s'$,
\i $b^0\in {\cal B}$ is an 
{\bf initial condition}, 
\i $D^0\subseteq {\cal E}$ is a set of 
{\bf disclosed terms},
values of these terms 
are known to both 
the process
$P$
and the environment
at the initial moment, and\i
$H^0\subseteq {\cal X}$ is a 
set of
{\bf hidden variables}.
\ei

A set of all processes
 is denoted by ${\cal P}$,
$\forall\,P\in {\cal P}$
the records
$S_P$, $s^0_P$, 
$R_P$, 
$b^0_P$, 
$D^0_P$, 
$H^0_P$
denote the corresponding components of $P$.
A set of variables 
occurring in $P$
is denoted by $X_P$. 
A process $P$ 
such that $R_P=\emptyset$ is denoted by 
{\bf 0}.

A  transition 
$s\ra{a}s'$ is said to be 
an {\bf input}, an
{\bf output},
or an {\bf internal transition}, 
if $a$ is an input, an output,
or an internal action, respectively.

A process $P$ can be represented
as a  graph (denoted by
the same symbol $P$): its nodes 
are states from 
$S_P$,  and edges are corresponded to
transitions from 
$R_P$: each transition
$s\ra{a}s'$ 
corresponds to an edge
from $s_1$ to $s_2$
labelled by $a$.
We  assume that for each process $P$
under consideration the graph $P$ is acyclic.

\subsection{An execution
of a process}

An execution of a process $P\in {\cal P}$ 
can be informally understood as a
walk on the graph $P$ starting from $s^0_P$, with 
an execution of actions that are 
labels of traversed edges. 
At each step $i\geq 0$
of this walk 
there are defined 
\bi
\i a state 
$s_i\in S_P$ of the process 
$P$ at the moment $i$,
\i a condition $b_i\in {\cal B}$
on variables of $P$ at the moment $i$, and
\i a set $D_i\subseteq {\cal E}$ of 
{\bf disclosed messages}
at the moment $i$,
i.e. 
messages known
to both 
the process $P$ 
and  the environment 
at the moment  $i$.
\ei

An {\bf execution}
of a process $P\in {\cal P}$ is 
 a sequence of the form
$$(s^0_P,b^0_P,D^0_P)=
(s_0,b_0,D_0)
\ra{a_1}
(s_1,b_1,D_1)
\ra{a_2}
\ldots
\ra{a_n}
(s_n,b_n,D_n)$$
where
$\forall\,i=1,\ldots, n\;\;(s_{i-1}\ra{a_{i}}s_{i})
\in R_P$, 
$(b_{i},D_{i})=
(b_{i-1},D_{i-1})a_{i}$,
and
$$\forall\,b\in {\cal B},
D\subseteq {\cal E},
a\in {\cal A}\quad
(b,D)a
=
\left\{
\by
\mbox{$(b, D
\cup\{e\})$, 
if $a=d?e$ or $d!e$, where
$d\in D^b$,}\\
\mbox{$(b\wedge a, D)$, 
if $a\in {\cal B}$,}\\
\mbox{undefined, otherwise.}
\ey
\right.$$

We assume that 
a value of each 
variable
$x\in H^0_P$ is unique
and unknown
to an environment of $P$
at the  initial moment 
of any 
execution of $P$.

A set of all
executions of $P$
can be represented by a 
labelled tree $T_P$,
where
\bi\i a root $t_P^0$
of the tree $T_P$
is labelled by the triple
$(s^0_P,
b^0_P, D^0_P)$, and 
\i if the set of  edges
of $P$
outgoing from $s^0_P$
is
$\{s^0_P\ra{a_i}s_i\mid i=1,\ldots, m\}$,
then
for each $i\in \{1,\ldots,m\}$, such that 
$\exists\,(b_i,D_i)=(b_P^0,D^0_P)a_i$,
\bi\i 
$T_P$ has an edge
of the form 
$t^0_P\ra{a_i}t_i$, and
\i
a subtree growing from $t_i$
is $T_{P_i}$,
where
$P_i=(S_P,s_i,R_P,
b_i,D_i,H^0_P\setminus 
D_i^{b_i})$.
\ei\ei


The set of nodes of $T_P$
is denoted by the same record
$T_P$. For each node
$t\in T_P$ the records
 $s_t$, 
$b_t$,  $D_t$
denote corresponding 
components of 
a label of $t$.

$\forall\,t,t'\in T_P$
the record $t\to t'$
means that either
$t=t'$, or there is 
a path in $T_P$
of the form 
$t=t_0\ra{a_1}t_1\ra{a_2}\ldots\ra{a_m}t_m=t'$,
where $a_1,\ldots, a_m\in 
{\cal B}$.

\subsection{Observational 
equivalence of processes}

In this section we introduce a concept of observational  equivalence of processes.
This concept has the following  sense: processes
$P_1$ and $P_2$ are
observationally equivalent
iff for any external observer 
(which can observe a behavior 
of $P_1$ and $P_2$ by sending
and receiving  messages) these processes 
are indistinguishable.

An example of a pair of 
observationally equivalent processes  is the pair
\be{dsfasdfas4453225}
P_i=
(\{s^0_{i},s_i\},s^0_{i},
\{
\diagrw{s^0_{i}&
\pright{c\,!\,k_i(e_i)}&s_i}\}, \top,
\{c\}, \{k_i\})\quad
(i=1,2).\ee
$P_1$ and $P_2$
send a unique message
via  channel $c$
and   then terminate.
Any process observing
an execution of $P_1$
and $P_2$ is unable
to distinguish them.

Processes $P_1, 
P_2\in {\cal P}$,
are said to be 
{\bf observationally 
equivalent} iff
there is a binary relation
$\mu \subseteq T_{P_1}\times T_{P_2}$ satisfying
the following conditions:
\bn
\i $\forall\,(t_1,t_2)\in \mu\;\;\exists\,\mu_{t_1,t_2}\in Sim\big((D_{t_1},{b_{t_1}}),(D_{t_2},{b_{t_2}})\big)$, 
\i $(t^0_{P_1}, 
t^0_{P_2})\in \mu$, 
$\forall\,(d_1,d_2)\in\mu_{t^0_{P_1},t^0_{P_2}}\;\exists\,d\in {\cal E}:
d_i=_{b^0_{P_i}}d\;\;(i=1,2)$,
\i 
$\forall\,(t_1,t_2)\in \mu$, 
for each edge 
$t_1\ra{a_1}t_1'$, 
$\exists\,t'_2\in T_{P_2}:
(t'_1, t'_2)\in \mu$,
$\mu_{t_1,t_2}
\subseteq 
\mu_{t'_1,t'_2}$, 
\bi
\i if $a_1=d_1\triangleright
e_1$ 
$(\triangleright\in \{?,!\})$,
then 
$\exists\,
t,t'\in T_{P_2}$:
$t_2\to t$,
$t'\to t_2'$,
and
$\exists\,d_2,e_2
$: $t\ra{d_2\triangleright
e_2}t'$,
 $(d_1,d_2)\in \mu_{t'_1,t'_2}$, 
 $(e_1,e_2)\in \mu_{t'_1,t'_2}$, 
\i if $a_1\in {\cal B}$, then
$t_2\to t'_2$,
\ei
\i a condition which is 
symmetric to  
condition 3: 
for each pair 
$(t_1,t_2)\in \mu$, 
and each edge 
$t_2\ra{a_2}t_2'$
there is a node 
$t'_1\in T_{P_1}$,
such that 
$(t'_1, t'_2)\in \mu$,
etc.
\en
\newpage
For example,
processes $P_i\;(i=1,2)$
from \re{dsfasdfas4453225}
are 
observationally 
equivalent, because in this
case $T_{P_i}$
has the form 
$\diagrw{(s^i_0,
\top,\{c\})&
\pright{c\,!\,k_i(e_i)}&
(s^i,
\top,\{c,k_i(e_i)\})}$, 
and the required
$\mu$ is 
$\{(s_1^0,s_1),
(s_2^0,s_2)\}$.

\subsection{Operations on
processes}
\label{operationsonprocesses}

In this section we define 
 operations on processes which can be used for a
construction of complex processes from simpler ones.

 \subsubsection{Prefix action}

$\forall\,a\in {\cal A},\;
\forall\,P\in {\cal P}\quad
\l{a}P$ is a process 
defined as follows:
$$\by
S_{\l{a}P}\eam
\{s\}\sqcup S_P,\quad 
s^0_{\l{a}P}\eam s,\quad
R_{\l{a}P}\eam 
\{s\ra{a}s_P^0\}\sqcup R_P,\\
b^0_{\l{a}P}\eam 
b^0_{P}$,
$D^0_{\l{a}P}\eam X_{a}\cup D^0_{P}$, 
 $H^0_{\l{a}P}\eam H^0_{P}.\ey$$

An execution of  $\l{a}P$ 
can be informally
understood as follows:
at first the action
$a$ is executed, then
$\l{a}P$  is executed
just like $P$.

\subsubsection{Choice}

$\forall\,P_1,P_2\in {\cal P}\quad
P_1+P_2$ is a process 
defined as follows:
all states of $P_1$,
that also 
belong to $S_{P_2}$,
are replaced by fresh states,
and 
$$\by
S_{P_1+P_2}\eam \{s\}
\sqcup S_{P_1}\sqcup S_{P_2},\quad
s^0_{P_1+P_2}\eam s,\\
R_{P_1+P_2}\eam 
R_{P_1}\sqcup R_{P_2}\sqcup
\{s\ra{a}s'\mid (s^0_{P_i}\ra{a}s')\in 
R_{P_i},\;i\in \{1,2\}\},\\
b^0_{P_1+P_2}\eam
b^0_{P_1}\wedge
b^0_{P_2},\;\;
D^0_{P_1+P_2}\eam D^0_{P_1}\cup D^0_{P_2},\;\;
H^0_{P_1+P_2}\eam H^0_{P_1}\cup H^0_{P_2}.\ey$$

An execution of  $P_1+P_2$
can be 
understood as follows:
at first it 
is selected (non-deterministically)
a process $P_i\in\{P_1,P_2\}$
which can execute its first action,
and then $P_1+P_2$
  is executed as the
  selected process.

\subsubsection{Parallel composition}

$\forall\,
P_1,P_2\in {\cal P}\;\;
(P_1,P_2)$ is a process 
defined as follows:
all variables in $X_{P_1}
\setminus D^0_{P_1}$,
that also belong to $X_{P_2}
\setminus 
D^0_{P_2}$,
are replaced by fresh variables, 
and 
\bi
\i $S_{(P_1,P_2)}\eam 
S_{P_1} \times
S_{P_2}$,
$s^0_{(P_1,P_2)}\eam
(s^0_{P_1},s^0_{P_2})$,
\i 
$R_{(P_1,P_2)}$ consists of the 
following transitions:
\bi
\i $(s_1,s_2)\ra{a}(s'_1,s_2)$,
where 
$(s_1\ra{a}s'_1)\in R_{P_1}$,
$s_2\in S_{P_2}$,
\i $(s_1,s_2)\ra{a}(s_1,s'_2)$,
where
$s_1\in S_{P_1}$,
$(s_2\ra{a}s'_2)\in R_{P_2}$,
\i $\diagrw{(s_1,s_2)&
\plongright{(d_1=d_2)\wedge(e_1=e_2)}&
(s'_1,s'_2)}$, where
$(s_i\ra{a_i}s'_i)\in R_{P_i}\;
(i=1,2)$, \\
$\{a_1,a_2\}=
\{d_1!e_1, d_2?e_2\}$
(such transition is said to be
{\bf diagonal}),
\ei
\i $b^0_{(P_1,P_2)}\eam
b^0_{P_1}\wedge
b^0_{P_2}$,
$D^0_{(P_1,P_2)}\eam D^0_{P_1}\cup D^0_{P_2}$,
$H^0_{(P_1,P_2)}\eam H^0_{P_1}\sqcup H^0_{P_2}$.
\ei

An execution of $(P_1,P_2)$
can be 
understood as 
undeterministic interleaving
 of executions of $P_1$
 and $P_2$: at each moment of 
 an execution of $(P_1,P_2)$
\bi
\i either 
one of $P_1,P_2$ executes an action, and another is in waiting,
   \i or 
one of $P_1,P_2$
sends a message,
and another receives this message.
\ei

A process 
$(\ldots(P_1,P_2),\ldots, P_n)$
is denoted by
$(P_1,\ldots, P_n)$.

\subsubsection{Replication}

$\forall\,P\in{\cal P}$
a {\bf replication} of 
$P$ is a process $P^\wedge$
that can be understood as 
infinite parallel composition
$(P,P,\ldots)$,
and is defined as follows.

$\forall\,i\geq 1$ let
$P_i$ be a process which 
is obtained from $P$
by renaming of  
variables:
$\forall\,x\in X_P\setminus
D^0_P$
each occurrence of 
$x$ in $P$
is replaced by the variable
$x_i$, such that
all the variables 
$x_i$ are fresh.
Components
of $P^\wedge$
have the following form:
\bi
\i $S_{P^\wedge}\eam
\{(s_1,s_2,\ldots)\mid
\forall\,i\geq 1\;\;s_i\in S_P\}$,
 $s^0_{P^\wedge}\eam
 (s^0_P,s^0_P,\ldots)$,
\i $\forall\,(s_1,\ldots)\in 
S_{P^\wedge}$,
$\forall\,i\geq 1$,
$\forall\, 
(s_i\ra{a}s)\in R_{P_i}
\quad
R_{P^\wedge}$ contains
the transitions
\bi
\i $(s_1,\ldots)\ra{a}\;(s_1,\ldots,s_{i-1},s,s_{i+1},\ldots)$,
and
\i $\!\diagrw{(s_1,\ldots)&
\plongright{(d_1=d_2)\wedge(e_1=e_2)}&
(s_1,\ldots,s_{i-1},s,s_{i+1},\ldots,s_{j-1},s',s_{j+1},\ldots)}$, where\\
$(s_j\ra{a'}s')\in R_{P_j}$ 
for some $j\neq i$,
and $\{a,a'\}=
\{d_1!e_1, d_2?e_2\}$,
\ei
\i $b^0_{P^\wedge}\eam
b^0_P$,
$D^0_{P^\wedge}\eam D^0_{P}$,
$H^0_{P^\wedge}\eam\bigsqcup_{i\geq 1}H^0_{P_i}$.
\ei

\subsubsection{Hiding}

$\forall\,P\in {\cal P},\;
\forall\,X\subseteq {\cal X}\quad
P_X\eam (S_P,s^0_P,
R_P,
b^0_P,
D^0_P\setminus X, 
H^0_P\cup X).$

If
$X=
\{x_1,\ldots, x_n\}$,
then $P_X$ is denoted by
$P_{x_1,\ldots, x_n}$.\\

\refstepcounter{theorem}
{\bf Theorem \arabic{theorem}\label{coco}.}
Observational congruence 
preserves operations of
prefix action, 
parallel composition, 
replication and hiding.
$\blackbox$

\subsection{A sufficient
condition of an observational equivalence}

Let $P\in {\cal P}$.
A {\bf labeling of states}
 of $P$ is a set 
$\{(b_s, D_s) \mid s\in S\}$,
such that \bi
\i $S\subseteq S_P$,
$\forall\,s\in S\;\;
b_s\in {\cal B}$
and $D_s\subseteq {\cal E}$,
 $s^0_{P}\in S$,
$b_{s^0_{P}}=b^0_{P}$, 
$D_{s^0_{P}}=D^0_{P}$,
\i for each transition
$(s\ra{a}s')\in R_P$,
if $s'\in S$ then $s\in S$, and in this case
\bi
\i if $a=d\triangleright
e$, where $\triangleright\in \{?,!\}$, then 
$d\in D_s^{b_s}$,
$b_s\leq b_{s'}$,
$D_s\cup\{e\}\subseteq 
D_{s'}^{b_{s'}}$,
\i if $a\in {\cal B}$, then
$b_s\wedge a\leq b_{s'}$
and 
$D_s\subseteq D_{s'}$.
\ei\ei

$\forall\,s,s'\in S_P$
the record $s\to s'$
means that either
$s=s'$, or there is 
a set of thansitions
of the form 
$s=s_0\ra{a_1}s_1\ra{a_2}\ldots\ra{a_m}s_m=s'$,
where $a_1,\ldots, a_m\in 
{\cal B}$.\\


\refstepcounter{theorem}
{\bf Theorem \arabic{theorem}\label{th1}}
(a sufficient
condition of an observational equivalence).

Let $P_1,P_2 \in {\cal P}$,
where $S_{P_1}\cap  S_{P_2}=\emptyset$.
Then $P_1\approx P_2$,
if there are a binary relation
$\mu\subseteq
S_{P_1}\times  S_{P_2}$ 
and  labelings
$\{(b_s, D_s) \mid s\in S_{P_1}\}$,
$\{(b_s, D_s) \mid s\in S_{P_2}\}$
of states of $P_1$ and $P_2$ respectively,
such that 
\bn
\i each pair $(s_1,s_2)\in \mu$ is associated with 
$\mu_{s_1s_2}\in Sim\big((D_{s_1},{b_{s_1}}), (D_{s_2},{b_{s_2}})\big)$, 
\i $(s^0_{P_1}, 
s^0_{P_2})\in \mu$, 
and each element of 
the set $\mu^0\eam
\mu_{s^0_{P_1}s^0_{P_2}}$ 
has the form 
$(x,x)$,
where $x\in D^0_{P_1}\cap 
D^0_{P_2}$,
\i for each pair 
$(s_1,s_2)\in \mu$, 
and each transition
$(s_1\ra{a_1}s_1')\in R_{P_1}$ 
there is a state 
$s'_2\in S_{P_2}$,
such that 
$(s'_1, s'_2)\in \mu$,
$\mu_{s_1s_2}
\subseteq 
\mu_{s'_1s'_2}$, and
\bi
\i if $a_1$ is  input or
output, then
$a_1=x\triangleright
e_1$, where  
$\triangleright\in \{?,!\}$,
 $(x,x)\in  \mu^0$,
$\exists\,
s,s'\in S_{P_2}$:
$s_2\to s$,
$s'\to s_2'$,
$\exists\,e_2
$: $s\ra{x\triangleright
e_2}s'$,
 $(e_1,e_2)\in \mu_{s'_1s'_2}$, 
\i if $a_1\in {\cal B}$, then
$s_2\to s'_2$,
\ei
\i a condition which is 
symmetric to  
condition 3: 
for each pair 
$(s_1,s_2)\in \mu$
and each transition 
$s_2\ra{a_2}s_2'$, 
$\exists\,s'_1\in S_{P_1}:
(s'_1, s'_2)\in \mu$,
etc. $\blackbox$
\en




\refstepcounter{theorem}
{\bf Theorem \arabic{theorem}\label{thred}.}
Let $P$ be a process, 
$\{(D_s, b_s) \mid s\in S\}$ be a labelling of $P$,
and 
$R_P$ has an edge 
$s\ra{a}s'$ such that 
$s,s'\in S$, and $a$
has the form 
$d?k(e)$, where
$D_{s}^{b_s}$
does not contain 
$k$ and any term of the 
form $k(e')$.
Then $P\approx P'$,
where  $P'$ is
obtained from $P$ 
by removing  
the above edge
and  all unreachable (from $s_P^0$)
states
which appear after 
removing the edge. 
$\blackbox$

\section{Security protocols}

A {\bf security protocol (SP)} is a process $P\in {\cal P}$ of the form 
$(P_1,\ldots, P_n)_X$,
where $P_1,\ldots, P_n$
are processes corresponding to {\bf agents} involved in the 
SP, and $X\subseteq {\cal X}$ is a {\bf shared secret} of the agents.
In this section we present 
an application of the proposed 
approach to 
description, specification of 
properties and 
verification of several 
examples of 
SPs, all of them 
are analogs of examples from 
\cite{82}.

\subsection{A message
passing
through 
a hidden channel}\label{dfsadfdsafafdsa}

First example is 
a  simplest SP for a 
message passing 
through 
a hidden channel. 
This SP consists of a 
sending of a message $x$
from an agent $a$ to 
an agent $b$
through a channel  named 
$c$ (where only $a$ and $b$
know the name $c$ of this channel),
$b$ receives the message 
and stores it in variable $y$,
then $b$ behaves like a process $P$.
This SP is represented by the
diagram
$$
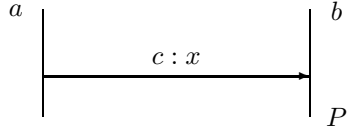
\begin{picture}(100,40)
\put(0,15){\vector(1,0){100}}
\put(0,0){\line(0,1){40}}
\put(100,0){\line(0,1){40}}
\put(-10,40){\makebox(0,0)[]{$a$}}
\put(110,40){\makebox(0,0)[]{$b$}}
\put(110,0){\makebox(0,0)[]{$P$}}
\put(50,22){\makebox(0,0)[]{$c:x$}}
\end{picture}
$$

A behavior of 
$a$ and $b$ is represented
by 
processes $A$ and $B$
respectively,
$A\eam \l{c\,!\,x}\,{\bf 0}$, $B\eam \l{c\,?y}\,P$
(where $c\not\in P$).
The SP is represented 
by the process
$Sys\eam(A,B)_c$.
Graph representations of 
processes in 
$Sys$ is the 
following:
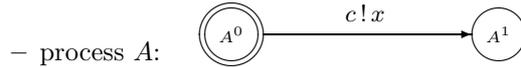
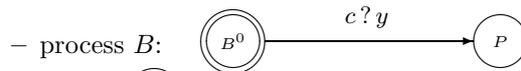
\bi
\i process $A$:
{\def\arraystretch{1}
{\small
$\qquad\by
\begin{picture}(110,0)

\put(0,0){\oval(20,20)}
\put(0,0){\oval(24,24)}
\put(0,0){\makebox(0,0)[c]{${\scriptstyle A^0}$}}

\put(100,0){\oval(20,20)}
\put(100,0){\makebox(0,0)[c]{${\scriptstyle A^1}$}}

\put(12,0){\vector(1,0){78}}

\put(50,5){\makebox(0,0)[b]{$c\,!\,x$}}

\end{picture}
\\
\vspace{3mm}
\ey
$
 }
}
\i process $B$:
{\def\arraystretch{1}
{\small
$\qquad\by
\begin{picture}(110,10)

\put(0,0){\oval(20,20)}
\put(0,0){\oval(24,24)}
\put(0,0){\makebox(0,0)[c]{${\scriptstyle B^0}$}}

\put(100,0){\oval(20,20)}
\put(100,0){\makebox(0,0)[c]{${\scriptstyle P}$}}

\put(12,0){\vector(1,0){78}}

\put(50,5){\makebox(0,0)[b]{$c\,?\,y$}}

\end{picture}
\\
\vspace{0mm}
\ey
$
 }
}\\
(where 
\begin{picture}(20,10)
\put(10,5){\oval(20,20)}
\put(10,5){\makebox(0,0)[c]{${\scriptstyle P}$}}
\end{picture}
denotes a subgraph corresponded
to the process $P$)
\vspace{3mm}
\i process $(A,B)$:\hspace{10mm}
{\def\arraystretch{1}
{\small
$\by
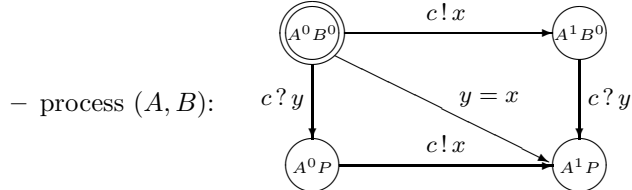
\begin{picture}(110,60)

\put(0,50){\oval(20,20)}
\put(0,50){\oval(24,24)}
\put(0,50){\makebox(0,0)[c]{${\scriptstyle A^0B^0}
$}}

\put(100,0){\oval(20,20)}
\put(100,0){\makebox(0,0)[c]{${\scriptstyle A^1P}
$}}

\put(100,50){\oval(20,20)}
\put(100,50){\makebox(0,0)[c]{${\scriptstyle A^1B^0}
$}}

\put(0,0){\oval(20,20)}
\put(0,0){\makebox(0,0)[c]{$
{\scriptstyle A^0P}$}}

\put(12,50){\vector(1,0){78}}
\put(50,55){\makebox(0,0)[b]{$c\,!\,x$}}

\put(10,0){\vector(1,0){80}}
\put(50,5){\makebox(0,0)[b]{$c\,!\,x$}}

\put(0,38){\vector(0,-1){28}}
\put(-3,25){\makebox(0,0)[r]{$c\,?\,y$}}

\put(100,40){\vector(0,-1){30}}
\put(103,25){\makebox(0,0)[l]{$c\,?\,y$}}

\put(8.5,41.5){\vector(2,-1){80}}
\put(55,25){\makebox(0,0)[l]{$y=x$}}

\end{picture}
\\
\vspace{0mm}
\ey
$
 }
}\\
(where 
\begin{picture}(20,10)
\put(10,5){\oval(20,20)}
\put(10,5){\makebox(0,0)[c]{${\scriptstyle A^0P}$}}
\end{picture}
and
\begin{picture}(20,10)
\put(10,5){\oval(20,20)}
\put(10,5){\makebox(0,0)[c]{$
{\scriptstyle A^1P}$}}
\end{picture}
denote subgraphs corresponded
to copies of $P$
(nodes of these graphs
are denoted by $A_is$,
where $i=0,1$, and
$s\in S_P$),
and the arrow from 
\begin{picture}(20,20)
\put(10,5){\oval(20,20)}
\put(10,5){\makebox(0,0)[c]{${\scriptstyle A^0P}$}}
\end{picture}
to
\begin{picture}(20,20)
\put(10,5){\oval(20,20)}
\put(10,5){\makebox(0,0)[c]{${\scriptstyle A^1P}$}}
\end{picture}
denotes a set of corresponding
transitions 
from $A^0s$ to
$A^1s$, where $s\in S_P$).
\vspace{3mm}
\ei

On the reason of 
theorem \ref{thred},
the process $(A,B)_c$
is observationally equivalent
to the process
{\def\arraystretch{1}
{\small
$\qquad\by
\begin{picture}(110,14)

\put(0,0){\oval(20,20)}
\put(0,0){\oval(24,24)}
\put(0,0){\makebox(0,0)[c]{${\scriptstyle A^0B^0}$}}

\put(100,0){\oval(20,20)}
\put(100,0){\makebox(0,0)[c]{${\scriptstyle A^1P}$}}

\put(12,0){\vector(1,0){78}}

\put(50,5){\makebox(0,0)[b]{$y=x$}}

\end{picture}
\\
\vspace{0mm}
\ey
$
 }
}.\\

The process model 
allows formally describe
and verify properties of 
integrity and secrecy 
of the above SP.
These properties are as follows.

\bi\i An {\bf integrity} of the SP 
is the following property: 
after a completion of the SP
agent $b$ receives 
the same message
that has been sent
 by agent $a$.   
   
\i A {\bf secrecy} of the SP 
is the following property: 
\bi
\i for each pair $x_1,x_2$ 
of messages, 
which $a$ can send  $b$ 
by this SP, and
\i for each 
two sessions  
of this SP, where 
the first session
is a passing of 
$x_1$, and the second one is
a passing of $x_2$,
\ei
any external 
(i.e. different from $a$ and $b$)
agent, observing an execution 
of these sessions, is unable 
to extract  from the 
observed information 
any knowledge about the messages $x_1$ and $x_2$: 
whether the messages are the same or different
(unless these
  knowledges are not disclosed
by participants $a$, $b$).

More accurately, the 
secrecy property can be 
described as follows:
for any pair $x_1,x_2$ of messages,
which $a$ can send 
$b$ by an execution of this SP
\bi
\i
if for any external observer
the process  $\l{y=x_1}\,P$
(which describes a behavior
of the agent $b$ after receiving
$x_1$)
is indistinguishable from 
the process $\l{y=x_2}\,P$
(which describes a behavior
of $b$ after receiving
$x_2$),
\i then for any sessions  of an 
execution of this SP, where the first one 
is a passing of $x_1$, 
and the second one 
is a passing of $x_2$, 
any external
agent, observing the execution of
these 
sessions, 
can not determine,
 are identical or different
messages 
transmitted in those sessions.
\ei
\ei

A formal description and verification of the 
properties of integrity and 
secrecy
of this SP is as follows.

\bn
\i  A property of 
{\bf integrity} 
is described by the proposition
   \be{auth}Sys\simplus \tilde{Sys}\ee
where  $\tilde{Sys}$ 
describes a SP which is 
defined like the original SP,
but with the following
   modification of 
     $b$:
after a receiving  
a message and storing it in a fresh variable $y'$,
a value of $y$ is 
changed on a value
that  $a$  
sent really.
A behavior of modified $b$ is described
   by  the process 
$\tilde B\eam
   \l{c\,?y'}\,\l{y=x}\,P$,
and the process
   $\tilde{Sys}$
   has the form $(A,\tilde B)_c$.
   
  Now we prove \re{auth}.
  The definition
  of operations on processes
   implies that
   \be{sdfasdfasdafrrrr}
   \by Sys\simplus\l{y=x}\,P,\quad
   \tilde{Sys}\simplus\l{y'=x}\,\l{y=x}\,P,\ey\ee
that implies
 \re{auth},
because $y'\not\in \l{y=x}\,P. \;\blackbox$

\i  A property of
 {\bf secrecy} of this SP
is described by the implication
\be{secr}
\by
\l{y=x_1}\,P\simplus \l{y=x_2}\,P\quad
\Rightarrow\quad
\l{x=x_1}\,
Sys \simplus \l{x=x_2}Sys
\\\mbox{(where
$x_1,x_2$ are fresh variables).}
\ey\ee

Now we prove \re{secr}.
The the premise of  implication
   \re{secr}
implies the statement
$$\l{y=x}\,
\l{y=x_1}\,P\simplus 
\l{y=x}\,
\l{y=x_2}\,P,$$
which is equivalent to
the statement
\be{sdfadsfasf}\l{x=x_1}\,
\l{y=x}\,P\simplus 
\l{x=x_2}\,
\l{y=x}\,P.\ee

\re{sdfadsfasf}
and
first proposition in
\re{sdfasdfasdafrrrr}
imply 
$$\l{x=x_1}\,Sys
\simplus
\l{x=x_1}\,\l{y=x}\,P
\simplus
\l{x=x_2}\,\l{y=x}\,P
\simplus
\l{x=x_2}\,Sys.\;\blackbox$$

\en

\subsection{A SP with 
a creation of a new channel}

Second SP consists of a  
message passing 
 from  $a$
to $b$, 
with an assumption that 
a channel for this passing
should be created during the execution of the SP.
An auxiliary  agent $t$
is used
in the SP 
($t$ is a trusted intermediary), and
it is assumed that a name of
a created channel 
must be known only to  $a$, $b$, and $t$.

This SP is represented by the diagram
$$
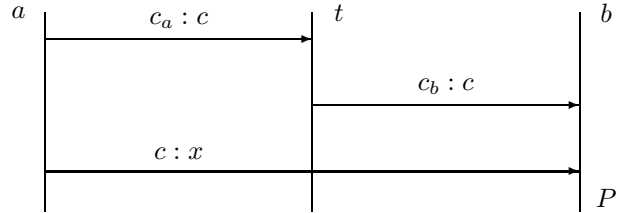
\begin{picture}(0,70)
\put(-100,55){\vector(1,0){100}}
\put(0,30){\vector(1,0){100}}
\put(-100,5){\vector(1,0){200}}
\put(0,-10){\line(0,1){75}}
\put(-100,-10){\line(0,1){75}}
\put(100,-10){\line(0,1){75}}
\put(-110,65){\makebox(0,0)[]{$a$}}
\put(10,65){\makebox(0,0)[]{$t$}}
\put(110,65){\makebox(0,0)[]{$b$}}
\put(110,-5){\makebox(0,0)[]{$P$}}
\put(-50,62){\makebox(0,0)[]{$c_a:c$}}
\put(50,37){\makebox(0,0)[]{$c_b:c$}}
\put(-50,12){\makebox(0,0)[]{$c:x$}}
\end{picture}
$$

A behavior of agents
$a,t,b$ 
is represented by the processes
$A$, $T$, $B$, where
$$
A\eam \l{c_a\,!\,c}\,\l{c\,!\,x}\,{\bf 0},\quad
T\eam \l{c_a\,?\,c}\,\l{c_b\,!\,c}\,{\bf 0},\quad
B\eam \l{c_b\,?\,c}\,\l{c\,?\,y}\,P.$$

The SP is represented 
by the process
$Sys\eam(A_c,T,B)_{c_a,c_b}$.

A formal description 
of integrity and secrecy of the SP is  represented by propositions
\re{auth} and \re{secr},
where
$\tilde{Sys}\eam(A_c,T,\tilde B)_{c_a,c_b}$,
$\tilde B\eam 
 \l{c_b\,?\,c}\,\l{c\,?\,y'}\,\l{y=x}\,P.
$


\subsection{A passing
of an encrypted message}
\label{dfsadfdsafafdsa1}

Third example is a
 SP, which involves agents $a$ and $b$ having a common key $k$
(only $a$ and
$b$ know 
$k$),  $a$ and $b$
can encrypt and decrypt  
messages by this key
using a symmetric encryption system.
The SP is as follows:
\bi
\i $a$ sends $b$
a ciphertext $k(x)$
through an open channel $c$,
\i $b$ receives the ciphertext,
 decrypts it,
 stores the extracted message
$x$ in the variable
$y$,  then behaves as a process
$P$.
\ei

This SP is represented by the diagram \hspace{5mm}
$
\begin{picture}(100,40)
\put(0,15){\vector(1,0){100}}
\put(0,0){\line(0,1){40}}
\put(100,0){\line(0,1){40}}
\put(-10,40){\makebox(0,0)[]{$a$}}
\put(110,40){\makebox(0,0)[]{$b$}}
\put(110,0){\makebox(0,0)[]{$P$}}
\put(50,22){\makebox(0,0)[]{$c:k(x)$}}
\end{picture}
$\hspace{7mm}.

A behavior of agents
 $a$ and
$b$ is represented by the 
processes
$A$ and $B$, where
$A\eam \l{c\,!\,k(x)}\,{\bf 0}$,
$B\eam \l{c\,?k(y)}\,P$, 
and
the SP is represented by $Sys\eam(A,B)_k$.


A formal description of the properties of integrity and secrecy of the SP is represented by 
\re{auth} and \re{secr},
where
$\tilde{Sys}\eam
(A,\tilde B)_{k}$, 
$\tilde B\eam 
 \l{c\,?\,k(y')}\,\l{y=x}\,P.$

An integrity  property of the SP is 
proposition \re{auth}, 
which in this case has the form
$\left(\l{c\,!\,k(x)}\,{\bf 0},
\,\l{c\,?k(y)}\,P\right)_k\simplus
\left(\l{c\,!\,k(x)}\,{\bf 0},
\,\l{c\,?k(y')}\,\l{y=x}\,P\right)_k$,
and can be proven with
use of theorem \ref{th1}.
To prove the secrecy 
property 
we prove implication \re{secr}.
With use of theorem \ref{th1}
it is not so difficult to prove that
\re{sdfasdfasdafrrrr}
and the premise of 
implication
   \re{secr}
   imply 
$Sys\simplus \l{y=x}\,P
\simplus  \l{y=x'}\,P,$
that proves \re{secr}. 

\subsection{
Wide-Mouth Frog
}

A SP
{\bf Wide-Mouth Frog (WMF)}
is intended for
a passing of an encrypted message $k(x)$
from an agent $a$
to an agent $b$
with use of a trusted
agent $t$, 
open channels
$c_a$, $c_b$, $c$,
and keys
$k_a$, $k_b$, $k$,
where $k_a$ should be 
known only to
$a$ and $t$,
$k_b$ should be known only to
$b$ and $t$, and 
$k$
should be known only to
$a$, $b$ and $t$.
This SP is represented by the 
diagram
$$
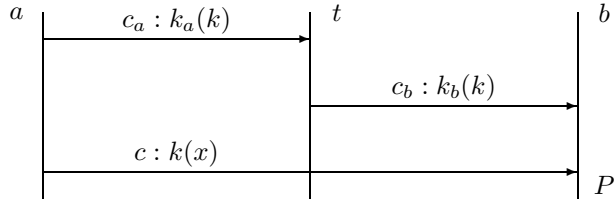
\begin{picture}(0,70)
\put(-100,55){\vector(1,0){100}}
\put(0,30){\vector(1,0){100}}
\put(-100,5){\vector(1,0){200}}
\put(0,-5){\line(0,1){70}}
\put(-100,-5){\line(0,1){70}}
\put(100,-5){\line(0,1){70}}
\put(-110,65){\makebox(0,0)[]{$a$}}
\put(10,65){\makebox(0,0)[]{$t$}}
\put(110,65){\makebox(0,0)[]{$b$}}
\put(110,0){\makebox(0,0)[]{$P$}}
\put(-50,62){\makebox(0,0)[]{$c_a:k_a(k)$}}
\put(50,37){\makebox(0,0)[]{$c_b:k_b(k)$}}
\put(-50,12){\makebox(0,0)[]{$c:k(x)$}}
\end{picture}
$$

A behavior  of agents
$a,t,b$ is represented 
by  processes $A,T,B$, where
\\$
A\eam \Big(\l{c_a\,!\,k_a(k)}\,
\l{c\,!\,k(x)}\,
{\bf 0}\Big)_k,\;
T\eam \l{c_a\,?\, k_a(k_T)}\,
\l{c_b\,!\,{k_b(k_T)}}\,{\bf 0}$, \\
$B\eam \l{c_b\,?\, k_b(k_B)}\,
\l{c\,?\, k_B(y)}\,P.$
The SP  is represented 
by 
$Sys\eam (A,T,B)_{k_a,k_b}$.

A formal description of the properties of integrity and secrecy of the SP is represented by the 
propositions
\re{auth} and \re{secr}, where
$$
\tilde{Sys}\eam(A,T,\tilde B)_{k_a,k_b},\quad
\tilde B\eam 
\l{c_b\,?\, k_b(k_B)}\,
\l{c\,?\, k_B(y')}\,
\l{y=x}\,P.
$$
Graph representations of 
processes involved in 
$Sys$ 
have the following form:
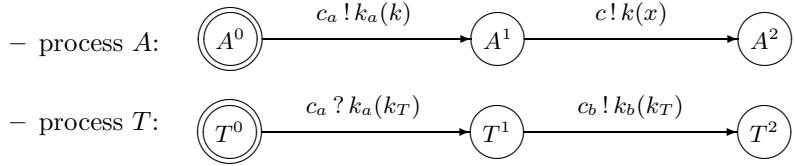
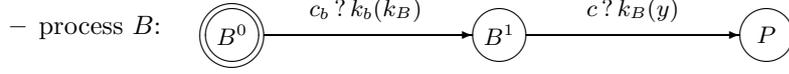
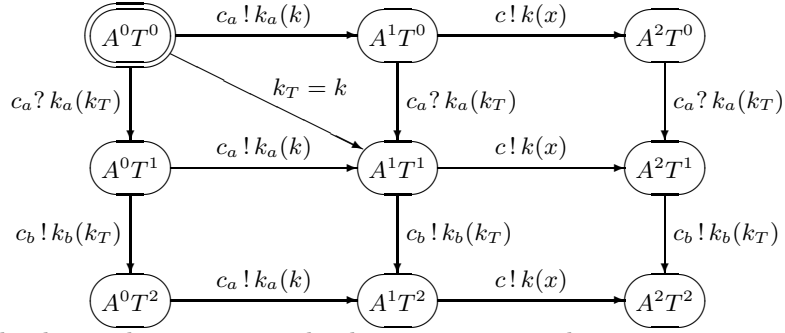
\bi\i process $A$:
{\def\arraystretch{1}
{\small
$\qquad\by
\begin{picture}(210,10)

\put(0,0){\oval(20,20)}
\put(0,0){\oval(24,24)}
\put(0,0){\makebox(0,0)[c]{$A^0$}}

\put(100,0){\oval(20,20)}
\put(100,0){\makebox(0,0)[c]{$A^1$}}

\put(200,0){\oval(20,20)}
\put(200,0){\makebox(0,0)[c]{$A^2$}}

\put(12,0){\vector(1,0){78}}
\put(110,0){\vector(1,0){80}}

\put(50,5){\makebox(0,0)[b]{$c_a\,!\,k_a(k)$}}
\put(150,5){\makebox(0,0)[b]{$c\,!\,k(x)$}}

\end{picture}
\\
\vspace{0mm}
\ey
$
 }
}
\i process $T$:
{\def\arraystretch{1}
{\small
$\qquad\by
\begin{picture}(210,20)

\put(0,0){\oval(20,20)}
\put(0,0){\oval(24,24)}
\put(0,0){\makebox(0,0)[c]{$T^0$}}

\put(100,0){\oval(20,20)}
\put(100,0){\makebox(0,0)[c]{$T^1$}}

\put(200,0){\oval(20,20)}
\put(200,0){\makebox(0,0)[c]{$T^2$}}

\put(12,0){\vector(1,0){78}}
\put(110,0){\vector(1,0){80}}

\put(50,5){\makebox(0,0)[b]{$c_a\,?\,k_a(k_T)$}}
\put(150,5){\makebox(0,0)[b]{$c_b\,!\,{k_b(k_T)}$}}

\end{picture}
\\
\vspace{0mm}
\ey
$
 } 
}
\i process $B$:
{\def\arraystretch{1}
{\small
$\qquad\by
\begin{picture}(210,20)

\put(0,0){\oval(20,20)}
\put(0,0){\oval(24,24)}
\put(0,0){\makebox(0,0)[c]{$B^0$}}

\put(100,0){\oval(20,20)}
\put(100,0){\makebox(0,0)[c]{$B^1$}}

\put(200,0){\oval(20,20)}
\put(200,0){\makebox(0,0)[c]{$P$}}

\put(12,0){\vector(1,0){78}}
\put(110,0){\vector(1,0){80}}

\put(50,5){\makebox(0,0)[b]{$c_b\,?\,k_b(k_B)$}}
\put(150,5){\makebox(0,0)[b]{$c\,?\,{k_B(y)}$}}

\end{picture}
\\
\vspace{0mm}
\ey
$
 } 
}
\i process $(A,T)$:
{\def\arraystretch{1}
{\small
$$\by
\begin{picture}(210,110)

\put(0,100){\oval(30,20)}
\put(0,100){\oval(34,24)}
\put(0,100){\makebox(0,0)[c]{$A^0T^0$}}

\put(0,50){\oval(30,20)}
\put(0,50){\makebox(0,0)[c]{$A^0T^1$}}

\put(0,0){\oval(30,20)}
\put(0,0){\makebox(0,0)[c]{$A^0T^2$}}

\put(100,100){\oval(30,20)}
\put(100,100){\makebox(0,0)[c]{$A^1T^0$}}

\put(100,50){\oval(30,20)}
\put(100,50){\makebox(0,0)[c]{$A^1T^1$}}

\put(100,0){\oval(30,20)}
\put(100,0){\makebox(0,0)[c]{$A^1T^2$}}

\put(200,100){\oval(30,20)}
\put(200,100){\makebox(0,0)[c]{$A^2T^0$}}

\put(200,50){\oval(30,20)}
\put(200,50){\makebox(0,0)[c]{$A^2T^1$}}

\put(200,0){\oval(30,20)}
\put(200,0){\makebox(0,0)[c]{$A^2T^2$}}

\put(0,75){\makebox(0,0)[r]{
$c_a?\,k_a(k_T)$
}}

\put(100,75){\makebox(0,0)[l]{
$c_a?\,k_a(k_T)$
}}

\put(200,75){\makebox(0,0)[l]{
$c_a?\,k_a(k_T)$
}}

\put(0,25){\makebox(0,0)[r]{
$c_b\,!\,k_b(k_T)$
}}

\put(100,25){\makebox(0,0)[l]{
$c_b\,!\,k_b(k_T)$
}}

\put(200,25){\makebox(0,0)[l]{
$c_b\,!\,k_b(k_T)$
}}

\put(50,103){\makebox(0,0)[b]{
$c_a\,!\,k_a(k)$
}}
\put(50,53){\makebox(0,0)[b]{
$c_a\,!\,k_a(k)$
}}
\put(50,3){\makebox(0,0)[b]{
$c_a\,!\,k_a( k)$
}}

\put(150,103){\makebox(0,0)[b]{
$c\,!\, k(x)$
}}

\put(150,53){\makebox(0,0)[b]{
$c\,!\, k(x)$
}}

\put(150,3){\makebox(0,0)[b]{
$c\,!\, k(x)$
}}

\put(50,80){\makebox(0,0)[l]{
$ k_T=k$
}}

\put(0,88){\vector(0,-1){28}}
\put(100,90){\vector(0,-1){30}}
\put(200,90){\vector(0,-1){30}}
\put(0,40){\vector(0,-1){30}}
\put(100,40){\vector(0,-1){30}}
\put(200,40){\vector(0,-1){30}}

\put(15,93){\vector(2,-1){72}}

\put(17,100){\vector(1,0){68}}
\put(15,50){\vector(1,0){70}}
\put(15,0){\vector(1,0){70}}

\put(115,100){\vector(1,0){70}}
\put(115,50){\vector(1,0){70}}
\put(115,0){\vector(1,0){70}}

\end{picture}
\ey
$$}}
(the diagonal transition 
in the diagram
corresponds 
to a joint execution of 
the action 
$c_a\,!\,k_a( k)$ of  $A$,
and
the action
$c_a?\,k_a(k_T)$ of $T$),
\i  process $(A,T,B)$:\\
\tikzstyle{block} = [draw, ellipse,fill=none, node distance=3cm,
  minimum height=2em]
\tikzstyle{line} = [draw, -latex']
\resizebox {11,5cm} {9cm} 
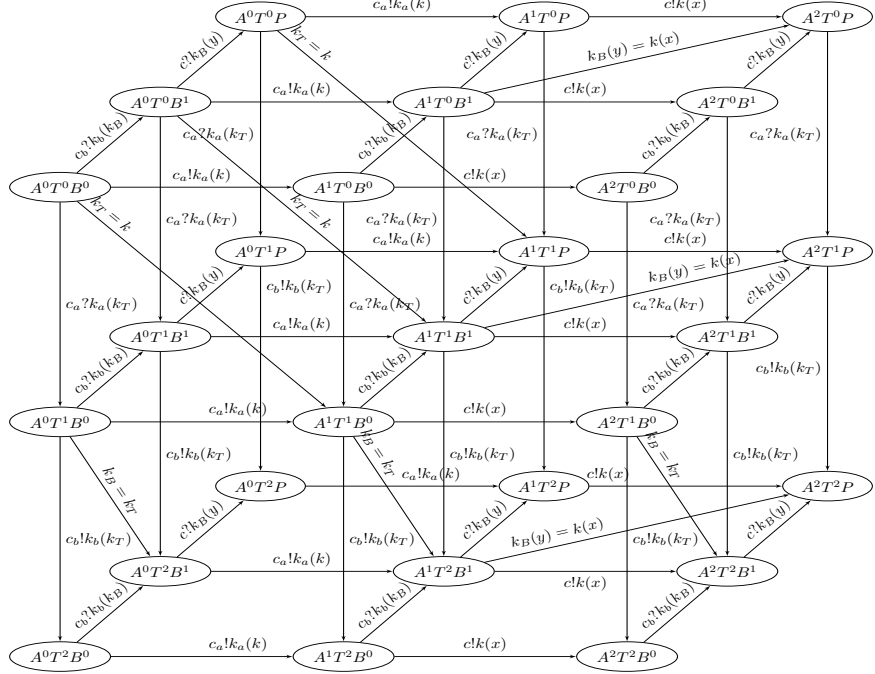
{
\begin{tikzpicture}	
	\node [block] (A^2T^0P) {$A^2T^0P$};
	\node [block, left of=A^2T^0P, node distance = 6cm] (A^1T^0P) {$A^1T^0P$};
	\node [block, left of=A^1T^0P, node distance = 6cm] (A^0T^0P) {$A^0T^0P$};

	\node [block, below left of=A^2T^0P] (A^2T^0B^1) {$A^2T^0B^1$};
	\node [block, left of=A^2T^0B^1, node distance = 6cm] (A^1T^0B^1) {$A^1T^0B^1$};
	\node [block, left of=A^1T^0B^1, node distance = 6cm] (A^0T^0B^1) {$A^0T^0B^1$};
	
	\node [block, below left of=A^2T^0B^1] (A^2T^0B^0) {$A^2T^0B^0$};
	\node [block, left of=A^2T^0B^0, node distance = 6cm] (A^1T^0B^0) {$A^1T^0B^0$};
	\node [block, left of=A^1T^0B^0, node distance = 6cm] (A^0T^0B^0) {$A^0T^0B^0$};

	\node [block, below of=A^2T^0P, node distance=2.3in] (A^2T^1P) {$A^2T^1P$};
	\node [block, left of=A^2T^1P, node distance = 6cm] (A^1T^1P) {$A^1T^1P$};
	\node [block, left of=A^1T^1P, node distance = 6cm] (A^0T^1P) {$A^0T^1P$};
	
	\node [block, below left of=A^2T^1P] (A^2T^1B^1) {$A^2T^1B^1$};
	\node [block, left of=A^2T^1B^1, node distance = 6cm] (A^1T^1B^1) {$A^1T^1B^1$};
	\node [block, left of=A^1T^1B^1, node distance = 6cm] (A^0T^1B^1) {$A^0T^1B^1$};
	
	\node [block, below left of=A^2T^1B^1] (A^2T^1B^0) {$A^2T^1B^0$};
	\node [block, left of=A^2T^1B^0, node distance = 6cm] (A^1T^1B^0) {$A^1T^1B^0$};
	\node [block, left of=A^1T^1B^0, node distance = 6cm] (A^0T^1B^0) {$A^0T^1B^0$};

	\node [block, below of=A^2T^1P, node distance=2.3in] (A^2T^2P) {$A^2T^2P$};
	\node [block, left of=A^2T^2P, node distance = 6cm] (A^1T^2P) {$A^1T^2P$};
	\node [block, left of=A^1T^2P, node distance = 6cm] (A^0T^2P) {$A^0T^2P$};
	
	\node [block, below left of=A^2T^2P] (A^2T^2B^1) {$A^2T^2B^1$};
	\node [block, left of=A^2T^2B^1, node distance = 6cm] (A^1T^2B^1) {$A^1T^2B^1$};
	\node [block, left of=A^1T^2B^1, node distance = 6cm] (A^0T^2B^1) {$A^0T^2B^1$};
	
	\node [block, below left of=A^2T^2B^1] (A^2T^2B^0) {$A^2T^2B^0$};
	\node [block, left of=A^2T^2B^0, node distance = 6cm] (A^1T^2B^0) {$A^1T^2B^0$};
	\node [block, left of=A^1T^2B^0, node distance = 6cm] (A^0T^2B^0) {$A^0T^2B^0$};
	
	
	\path [line] (A^0T^0P) -- node[auto] {$c_a!k_a(k)$} (A^1T^0P);
	\path [line] (A^1T^0P) -- node[auto] {$c!k(x)$} (A^2T^0P);
	\path [line] (A^0T^0B^1) -- node[sloped, above] {$c?k_B(y)$} (A^0T^0P);
	\path [line] (A^0T^0B^1) -- node[pos=0.5, above] {$c_a!k_a(k)$} (A^1T^0B^1);
	\path [line] (A^1T^0B^1) -- node[auto] {$c!k(x)$} (A^2T^0B^1);
	\path [line] (A^1T^0B^1) -- node[sloped, above] {$c?k_B(y)$} (A^1T^0P);
	\path [line] (A^2T^0B^1) -- node[sloped, above] {$c?k_B(y)$} (A^2T^0P);
	\path [line] (A^0T^0B^0) -- node[sloped, above] {$c_b?k_b(k_B)$} (A^0T^0B^1);
	\path [line] (A^0T^0B^0) -- node[pos=0.5, above] {$c_a!k_a(k)$} (A^1T^0B^0);
	\path [line] (A^1T^0B^0) -- node[sloped, above] {$c_b?k_b(k_B)$} (A^1T^0B^1);
	\path [line] (A^1T^0B^0) -- node[auto] {$c!k(x)$} (A^2T^0B^0);
	\path [line] (A^2T^0B^0) -- node[sloped, above] {$c_b?k_b(k_B)$} (A^2T^0B^1);
	
	\path [line] (A^1T^0B^1) -- node[pos=.5, sloped, above] {$k_B(y)=k(x)$} (A^2T^0P);
	
	\path [line] (A^0T^1P) -- node[auto] {$c_a!k_a(k)$} (A^1T^1P);
	\path [line] (A^1T^1P) -- node[auto] {$c!k(x)$} (A^2T^1P);
	\path [line] (A^0T^1B^1) -- node[sloped, above] {$c?k_B(y)$} (A^0T^1P);
	\path [line] (A^0T^1B^1) -- node[auto] {$c_a!k_a(k)$} (A^1T^1B^1);
	\path [line] (A^1T^1B^1) -- node[auto] {$c!k(x)$} (A^2T^1B^1);
	\path [line] (A^1T^1B^1) -- node[sloped, above] {$c?k_B(y)$} (A^1T^1P);
	\path [line] (A^2T^1B^1) -- node[sloped, above] {$c?k_B(y)$} (A^2T^1P);
	\path [line] (A^0T^1B^0) -- node[sloped, above] {$c_b?k_b(k_B)$} (A^0T^1B^1);
	\path [line] (A^0T^1B^0) -- node[pos=0.7, above] {$c_a!k_a(k)$} (A^1T^1B^0);
	\path [line] (A^1T^1B^0) -- node[sloped, above] {$c_b?k_b(k_B)$} (A^1T^1B^1);
	\path [line] (A^1T^1B^0) -- node[auto] {$c!k(x)$} (A^2T^1B^0);
	\path [line] (A^2T^1B^0) -- node[sloped, above] {$c_b?k_b(k_B)$} (A^2T^1B^1);
	
	\path [line] (A^1T^1B^1) -- node[pos=.7, sloped, above] {$k_B(y)=k(x)$} (A^2T^1P);
	
	\path [line] (A^0T^2P) -- node[pos=0.65, above] {$c_a!k_a(k)$} (A^1T^2P);
	\path [line] (A^1T^2P) -- node[pos= 0.1,above] {$c!k(x)$} (A^2T^2P);
	\path [line] (A^0T^2B^1) -- node[sloped, above] {$c?k_B(y)$} (A^0T^2P);
	\path [line] (A^0T^2B^1) -- node[auto] {$c_a!k_a(k)$} (A^1T^2B^1);
	\path [line] (A^1T^2B^1) -- node[below] {$c!k(x)$} (A^2T^2B^1);
	\path [line] (A^1T^2B^1) -- node[sloped, above] {$c?k_B(y)$} (A^1T^2P);
	\path [line] (A^2T^2B^1) -- node[sloped, above] {$c?k_B(y)$} (A^2T^2P);
	\path [line] (A^0T^2B^0) -- node[sloped, above] {$c_b?k_b(k_B)$} (A^0T^2B^1);
	\path [line] (A^0T^2B^0) -- node[pos=0.7, above] {$c_a!k_a(k)$} (A^1T^2B^0);
	\path [line] (A^1T^2B^0) -- node[sloped, above] {$c_b?k_b(k_B)$} (A^1T^2B^1);
	\path [line] (A^1T^2B^0) -- node[auto] {$c!k(x)$} (A^2T^2B^0);
	\path [line] (A^2T^2B^0) -- node[sloped, above] {$c_b?k_b(k_B)$} (A^2T^2B^1);
	
	\path [line] (A^1T^2B^1) -- node[pos=.25, sloped, above] {$k_B(y)=k(x)$} (A^2T^2P);

	\path [line] (A^0T^0P) 	 -- node[pos=0.5, left]  {$c_a?k_a(k_T)$} (A^0T^1P);
	\path [line] (A^0T^0B^1)  -- node[pos=0.5, right]  {$c_a?k_a(k_T)$} (A^0T^1B^1);
	\path [line] (A^0T^0B^0) 	 -- node[pos=0.5, right] {$c_a?k_a(k_T)$} (A^0T^1B^0);
	\path [line] (A^0T^1P)  -- node[pos=0.1, right] {$c_b!k_b(k_T)$} (A^0T^2P);
	\path [line] (A^0T^1B^1) -- node[auto]           {$c_b!k_b(k_T)$} (A^0T^2B^1);
	\path [line] (A^0T^1B^0)  -- node[auto]           {$c_b!k_b(k_T)$} (A^0T^2B^0);
	
	\path [line] (A^0T^1B^0) -- node[sloped, above] {$k_B=k_T$} (A^0T^2B^1);

	\path [line] (A^1T^0P)   -- node[pos=0.5, left]  {$c_a?k_a(k_T)$} (A^1T^1P);
	\path [line] (A^1T^0B^1)  -- node[pos=0.5, left]  {$c_a?k_a(k_T)$} (A^1T^1B^1);
	\path [line] (A^1T^0B^0)   -- node[pos=0.5, right] {$c_a?k_a(k_T)$} (A^1T^1B^0);
	\path [line] (A^1T^1P)  -- node[pos=0.1, right] {$c_b!k_b(k_T)$} (A^1T^2P);
	\path [line] (A^1T^1B^1) -- node[auto]           {$c_b!k_b(k_T)$} (A^1T^2B^1);
	\path [line] (A^1T^1B^0)  -- node[auto]           {$c_b!k_b(k_T)$} (A^1T^2B^0);
	
	\path [line] (A^1T^1B^0)  -- node[pos=0.18, sloped, above]           {$k_B=k_T$} (A^1T^2B^1);
	
	\path [line] (A^2T^0P)   -- node[pos=0.5, left] {$c_a?k_a(k_T)$} (A^2T^1P);
	\path [line] (A^2T^0B^1)  -- node[pos=0.5, left] {$c_a?k_a(k_T)$} (A^2T^1B^1);
	\path [line] (A^2T^0B^0)   -- node[pos=0.5, right] {$c_a?k_a(k_T)$} (A^2T^1B^0);
	\path [line] (A^2T^1P)  -- node[pos=0.5, left] {$c_b!k_b(k_T)$} 	 (A^2T^2P);
	\path [line] (A^2T^1B^1) -- node[auto] {$c_b!k_b(k_T)$}     (A^2T^2B^1);
	\path [line] (A^2T^1B^0)  -- node[auto] {$c_b!k_b(k_T)$}     (A^2T^2B^0);
	
	\path [line] (A^2T^1B^0)  -- node[pos=0.2, sloped, above] {$k_B=k_T$} (A^2T^2B^1);

	\path [line] (A^0T^0P) -- node[pos=0.1, sloped, above] {$k_T=k$} (A^1T^1P);
	\path [line] (A^0T^0B^1) -- node[pos=0.5, sloped, above] {$k_T=k$} (A^1T^1B^1);
	\path [line] (A^0T^0B^0) -- node[pos=0.1, sloped, above] {$k_T=k$} (A^1T^1B^0);

\end{tikzpicture}
}

this diagram 
has  
diagonal transitions,
related to a joint execution of 
\bi
\i 
 action 
$c_b\,!\,k_b(k_T)$ of  $(A,T)$,
and
 action
$c_b?\,k_b(k_B)$ of $B$
(transitions of the form
$A^0T^1B^0\to A^0T^2B^1$,
$A^1T^1B^0\to A^1T^2B^1$,
$A^2T^1B^0\to A^2T^2B^1$,
labelled by $k_B=k_T$),
and
\i 
 action 
$c\,!\,k(x)$ of  $(A,T)$,
and
 action
$c?\,k_B(y)$ of $B$
(transitions of the form
$A^1T^0B^1\to A^2T^0P$,
$A^1T^1B^1\to A^2T^1P$,
$A^1T^2B^1\to A^2T^2P$,
labelled by $k_B(y)=k(x)$),
\ei
\ei

Process $Sys\eam
(A,T,B)_{k_a,k_b}$
has the same graph representation
as the above process
$(A,T,B)$. Its initial state
is $A^0T^0B^0$.


First reduction
of $Sys$
is based on an 
applying of theorem 
\ref{thred} for the cases
\bi
\i the edge is 
$\diagrw{A^0T^0B^0&\pright{c_a?k_a(k_T)}&A^0T^1B^0}$, 
$D_{A^0T^0B^0}=
\{c_a,c_b,c\}$, 
and 
\i the edge is 
$\diagrw{A^0T^0B^0&\pright{c_b?k_b(k_B)}&A^0T^0B^1}$, 
$D_{A^0T^0B^0}=\{c_a,c_b,c\}$.
\ei
Removing the above
edges 
and all nodes and edges which 
become
unreachable from $A^0T^0B^0$
will result the graph
\\
\tikzstyle{block} = [draw, ellipse,fill=none, node distance=3cm,
  minimum height=2em]
\tikzstyle{line} = [draw, -latex']
\resizebox {11,5cm} {9cm} 
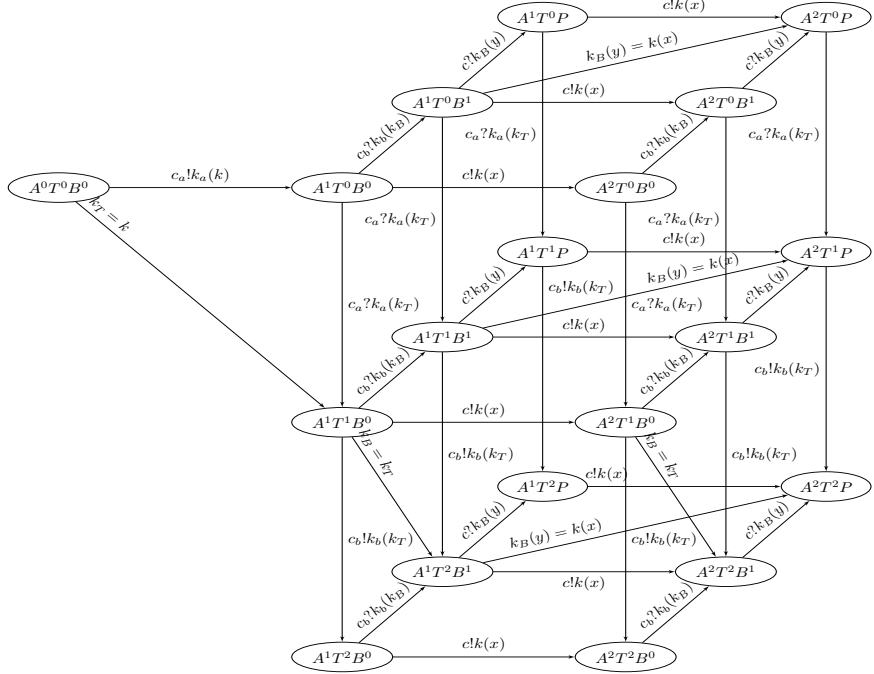
{
\begin{tikzpicture}	
	\node [block] (A^2T^0P) {$A^2T^0P$};
	\node [block, left of=A^2T^0P, node distance = 6cm] (A^1T^0P) {$A^1T^0P$};

	\node [block, below left of=A^2T^0P] (A^2T^0B^1) {$A^2T^0B^1$};
	\node [block, left of=A^2T^0B^1, node distance = 6cm] (A^1T^0B^1) {$A^1T^0B^1$};
	
	\node [block, below left of=A^2T^0B^1] (A^2T^0B^0) {$A^2T^0B^0$};
	\node [block, left of=A^2T^0B^0, node distance = 6cm] (A^1T^0B^0) {$A^1T^0B^0$};
	\node [block, left of=A^1T^0B^0, node distance = 6cm] (A^0T^0B^0) {$A^0T^0B^0$};

	\node [block, below of=A^2T^0P, node distance=2.3in] (A^2T^1P) {$A^2T^1P$};
	\node [block, left of=A^2T^1P, node distance = 6cm] (A^1T^1P) {$A^1T^1P$};
	
	\node [block, below left of=A^2T^1P] (A^2T^1B^1) {$A^2T^1B^1$};
	\node [block, left of=A^2T^1B^1, node distance = 6cm] (A^1T^1B^1) {$A^1T^1B^1$};
	
	\node [block, below left of=A^2T^1B^1] (A^2T^1B^0) {$A^2T^1B^0$};
	\node [block, left of=A^2T^1B^0, node distance = 6cm] (A^1T^1B^0) {$A^1T^1B^0$};

	\node [block, below of=A^2T^1P, node distance=2.3in] (A^2T^2P) {$A^2T^2P$};
	\node [block, left of=A^2T^2P, node distance = 6cm] (A^1T^2P) {$A^1T^2P$};
	
	\node [block, below left of=A^2T^2P] (A^2T^2B^1) {$A^2T^2B^1$};
	\node [block, left of=A^2T^2B^1, node distance = 6cm] (A^1T^2B^1) {$A^1T^2B^1$};
	
	\node [block, below left of=A^2T^2B^1] (A^2T^2B^0) {$A^2T^2B^0$};
	\node [block, left of=A^2T^2B^0, node distance = 6cm] (A^1T^2B^0) {$A^1T^2B^0$};
	
	
	\path [line] (A^1T^0P) -- node[auto] {$c!k(x)$} (A^2T^0P);
	\path [line] (A^1T^0B^1) -- node[auto] {$c!k(x)$} (A^2T^0B^1);
	\path [line] (A^1T^0B^1) -- node[sloped, above] {$c?k_B(y)$} (A^1T^0P);
	\path [line] (A^2T^0B^1) -- node[sloped, above] {$c?k_B(y)$} (A^2T^0P);
	\path [line] (A^0T^0B^0) -- node[pos=0.5, above] {$c_a!k_a(k)$} (A^1T^0B^0);
	\path [line] (A^1T^0B^0) -- node[sloped, above] {$c_b?k_b(k_B)$} (A^1T^0B^1);
	\path [line] (A^1T^0B^0) -- node[auto] {$c!k(x)$} (A^2T^0B^0);
	\path [line] (A^2T^0B^0) -- node[sloped, above] {$c_b?k_b(k_B)$} (A^2T^0B^1);
	
	\path [line] (A^1T^0B^1) -- node[pos=.5, sloped, above] {$k_B(y)=k(x)$} (A^2T^0P);
	
	\path [line] (A^1T^1P) -- node[auto] {$c!k(x)$} (A^2T^1P);
	\path [line] (A^1T^1B^1) -- node[auto] {$c!k(x)$} (A^2T^1B^1);
	\path [line] (A^1T^1B^1) -- node[sloped, above] {$c?k_B(y)$} (A^1T^1P);
	\path [line] (A^2T^1B^1) -- node[sloped, above] {$c?k_B(y)$} (A^2T^1P);
	\path [line] (A^1T^1B^0) -- node[sloped, above] {$c_b?k_b(k_B)$} (A^1T^1B^1);
	\path [line] (A^1T^1B^0) -- node[auto] {$c!k(x)$} (A^2T^1B^0);
	\path [line] (A^2T^1B^0) -- node[sloped, above] {$c_b?k_b(k_B)$} (A^2T^1B^1);
	
	\path [line] (A^1T^1B^1) -- node[pos=.7, sloped, above] {$k_B(y)=k(x)$} (A^2T^1P);
	
	\path [line] (A^1T^2P) -- node[pos= 0.1,above] {$c!k(x)$} (A^2T^2P);
	\path [line] (A^1T^2B^1) -- node[below] {$c!k(x)$} (A^2T^2B^1);
	\path [line] (A^1T^2B^1) -- node[sloped, above] {$c?k_B(y)$} (A^1T^2P);
	\path [line] (A^2T^2B^1) -- node[sloped, above] {$c?k_B(y)$} (A^2T^2P);
	\path [line] (A^1T^2B^0) -- node[sloped, above] {$c_b?k_b(k_B)$} (A^1T^2B^1);
	\path [line] (A^1T^2B^0) -- node[auto] {$c!k(x)$} (A^2T^2B^0);
	\path [line] (A^2T^2B^0) -- node[sloped, above] {$c_b?k_b(k_B)$} (A^2T^2B^1);
	
	\path [line] (A^1T^2B^1) -- node[pos=.25, sloped, above] {$k_B(y)=k(x)$} (A^2T^2P);


	\path [line] (A^1T^0P)   -- node[pos=0.5, left]  {$c_a?k_a(k_T)$} (A^1T^1P);
	\path [line] (A^1T^0B^1)  -- node[pos=0.5, left]  {$c_a?k_a(k_T)$} (A^1T^1B^1);
	\path [line] (A^1T^0B^0)   -- node[pos=0.5, right] {$c_a?k_a(k_T)$} (A^1T^1B^0);
	\path [line] (A^1T^1P)  -- node[pos=0.1, right] {$c_b!k_b(k_T)$} (A^1T^2P);
	\path [line] (A^1T^1B^1) -- node[auto]           {$c_b!k_b(k_T)$} (A^1T^2B^1);
	\path [line] (A^1T^1B^0)  -- node[auto]           {$c_b!k_b(k_T)$} (A^1T^2B^0);
	
	\path [line] (A^1T^1B^0)  -- node[pos=0.18, sloped, above]           {$k_B=k_T$} (A^1T^2B^1);
	
	\path [line] (A^2T^0P)   -- node[pos=0.5, left] {$c_a?k_a(k_T)$} (A^2T^1P);
	\path [line] (A^2T^0B^1)  -- node[pos=0.5, left] {$c_a?k_a(k_T)$} (A^2T^1B^1);
	\path [line] (A^2T^0B^0)   -- node[pos=0.5, right] {$c_a?k_a(k_T)$} (A^2T^1B^0);
	\path [line] (A^2T^1P)  -- node[pos=0.5, left] {$c_b!k_b(k_T)$} 	 (A^2T^2P);
	\path [line] (A^2T^1B^1) -- node[auto] {$c_b!k_b(k_T)$}     (A^2T^2B^1);
	\path [line] (A^2T^1B^0)  -- node[auto] {$c_b!k_b(k_T)$}     (A^2T^2B^0);
	
	\path [line] (A^2T^1B^0)  -- node[pos=0.2, sloped, above] {$k_B=k_T$} (A^2T^2B^1);

	\path [line] (A^0T^0B^0) -- node[pos=0.1, sloped, above] {$k_T=k$} (A^1T^1B^0);

\end{tikzpicture}
}


This graph
also can be reduced
with use of theorem 
\ref{thred} for the 
following cases:
\bi
\i the edge is 
$\diagrw{A^1T^0B^0&\pright{c_b?k_b(k_B)}&A^1T^0B^1}$, 
$D_{A^1T^0B^0}=\{c_a,c_b,c,k_a(k)\}$,
\i  the edge is 
$\diagrw{A^1T^1B^0&\pright{c_b?k_b(k_B)}&A^1T^1B^1}$,
$D_{A^1T^1B^0}=\{c_a,c_b,c,k_a(k)\}$,
\i  the edge is 
$\diagrw{A^2T^0B^0&\pright{c_b?k_b(k_B)}&\!A^2T^0B^1}$,
$D_{A^2T^0B^0}=\{c_a,c_b,c,k_a(k), k(x)\}$,
\i  the edge is 
$\diagrw{A^2T^1B^0&\pright{c_b?k_b(k_B)}&\!A^2T^1B^1}$,
$D_{A^2T^1B^0}=\{c_a,c_b,c,k_a(k), k(x)\}$.
\ei

Removing
the above edges 
and all nodes and edges which 
become
unreachable from $A^0T^0B^0$
will result to the graph

{\def\arraystretch{1}
{\small
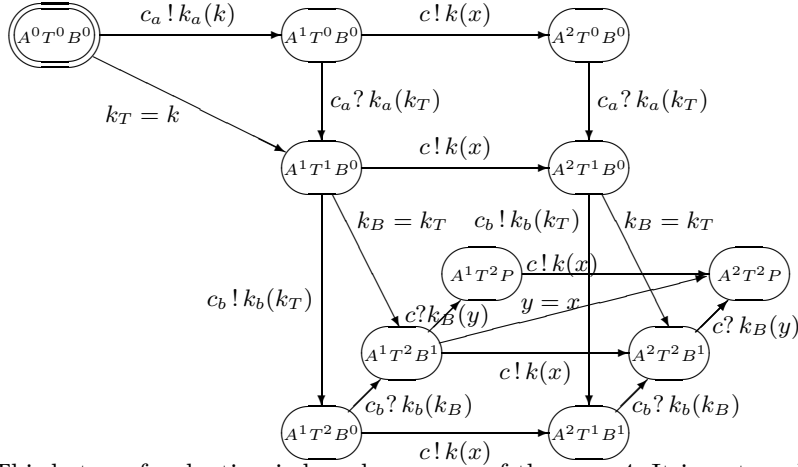
$$\by
\begin{picture}(270,150)

\put(0,150){\oval(30,20)}
\put(0,150){\oval(34,24)}
\put(0,150){\makebox(0,0)[c]{${\scriptstyle A^0T^0B^0}$}}

\put(100,150){\oval(30,20)}
\put(100,150){\makebox(0,0)[c]{${\scriptstyle A^1T^0B^0}$}}

\put(100,100){\oval(30,20)}
\put(100,100){\makebox(0,0)[c]{${\scriptstyle A^1T^1B^0}$}}

\put(100,0){\oval(30,20)}
\put(100,0){\makebox(0,0)[c]{${\scriptstyle A^1T^2B^0}$}}

\put(200,150){\oval(30,20)}
\put(200,150){\makebox(0,0)[c]{${\scriptstyle A^2T^0B^0}$}}

\put(200,100){\oval(30,20)}
\put(200,100){\makebox(0,0)[c]{${\scriptstyle A^2T^1B^0}$}}

\put(200,0){\oval(30,20)}
\put(200,0){\makebox(0,0)[c]{${\scriptstyle A^2T^1B^1}$}}

\put(100,125){\makebox(0,0)[l]{
$c_a?\, k_a(k_T)$
}}

\put(200,125){\makebox(0,0)[l]{
$c_a?\, k_a(k_T)$
}}

\put(100,50){\makebox(0,0)[r]{
$c_b\,!\, k_b(k_T)$
}}

\put(200,80){\makebox(0,0)[r]{
$c_b\,!\, k_b(k_T)$
}}

\put(50,153){\makebox(0,0)[b]{
$c_a\,!\, k_a( k)$
}}

\put(150,153){\makebox(0,0)[b]{
$c\,!\, k(x)$
}}

\put(150,103){\makebox(0,0)[b]{
$c\,!\, k(x)$
}}

\put(150,-3){\makebox(0,0)[t]{
$c\,!\, k(x)$
}}

\put(50,120){\makebox(0,0)[r]{
$ k_T=k$
}}

\put(100,140){\vector(0,-1){30}}
\put(200,140){\vector(0,-1){30}}

\put(100,90){\vector(0,-1){80}}
\put(200,90){\vector(0,-1){80}}

\put(14.5,142){\vector(2,-1){72}}

\put(17,150){\vector(1,0){68}}

\put(115,150){\vector(1,0){70}}
\put(115,100){\vector(1,0){70}}
\put(115,0){\vector(1,0){70}}

\put(130,30){\oval(30,20)}
\put(130,30){\makebox(0,0)[c]{${\scriptstyle A^1T^2B^1}$}}

\put(160,60){\oval(30,20)}
\put(160,60){\makebox(0,0)[c]{${\scriptstyle A^1T^2P}$}}

\put(230,30){\oval(30,20)}
\put(230,30){\makebox(0,0)[c]{${\scriptstyle A^2T^2B^1}$}}

\put(180,27){\makebox(0,0)[t]{
$c\,!\, k(x)$
}}

\put(190,68){\makebox(0,0)[t]{
$c\,!\, k(x)$
}}

\put(145,30){\vector(1,0){70}}

\put(175,60){\vector(1,0){70}}

\put(260,60){\oval(30,20)}
\put(260,60){\makebox(0,0)[c]{${\scriptstyle A^2T^2P}$}}

\put(110,8){\vector(1,1){12}}

\put(140,38){\vector(1,1){12}}
\put(210,8){\vector(1,1){12}}
\put(240,38){\vector(1,1){12}}
\put(240,38){\vector(1,1){12}}

\put(205,90){\vector(1,-2){25}}
\put(104,90){\vector(1,-2){25}}

\put(144,34){\vector(4,1){100}}

\put(113,10){\makebox(0,0)[l]{
$c_b?\, k_b(k_B)$
}}

\put(128,44){\makebox(0,0)[l]{
$c?k_B(y)$
}}

\put(213,10){\makebox(0,0)[l]{
$c_b?\, k_b(k_B)$
}}
\put(243,40){\makebox(0,0)[l]{
$c?\,k_B(y)$
}}

\put(110,80){\makebox(0,0)[l]{
$k_B=k_T$
}}
\put(210,80){\makebox(0,0)[l]{
$k_B=k_T$
}}

\put(200,48){\makebox(0,0)[r]{
$y=x$
}}

\end{picture}
\ey
$$
 }
}

Third step of reduction
is based on a use of
theorem \ref{thred}.
It is not so difficult that
there is a labelling 
for the process 
presented by the 
above graph:\\
$\left\{\by
D_{A^0T^0B^0}=\{c_a,c_b,c\},\\
D_{A^1T^0B^0}=D_{A^1T^1B^0}=
\{c_a,c_b,c, k_a(k)\}\\
D_{A^2T^0B^0}=D_{A^2T^1B^0}=
\{c_a,c_b,c, k_a(k), k(x)\}\\
D_{A^1T^2B^0}=
D_{A^1T^2B^1}=
D_{A^1T^2B^2}=
\{c_a, c_b, c,k_a(k), k_b(k_T)\},\\
D_{A^2T^2B^0}=
D_{A^2T^2B^1}=
D_{A^2T^2P}=
\{c_a, c_b, c,k_a(k), k_b(k_T),
k(x)\},\ey\right.
$\\
$\left\{\by
b_{A^0T^0B^0}=
b_{A^1T^0B^0}=
b_{A^2T^0B^0}=\top,\\
b_{A^1T^1B^0}=
b_{A^2T^1B^0}=b_{A^1T^2B^0}=
b_{A^2T^2B^0}=
(k=k_T),\\
b_{A^1T^2B^1}=
b_{A^2T^2B^1}=(k=k_T)\wedge
(k_T=k_B),\\
b_{A^2T^2P}=
(k=k_T)\wedge
(k_T=k_B)\wedge(x=y).
\ey\right.$\\
On the reason of 
theorem \ref{thred},
the edge 
$\diagrw{A^1T^2B^1&\pright{c?k_B(y)}&A^1T^2P}$ in the last diagram
can be removed, because
$b_{A^1T^2B^1}\leq (k_B=k)$,
and $(k\in D_{A^1T^2B^1})=\bot$.

The result of such removing is
the process below:
{\def\arraystretch{1}
{\small
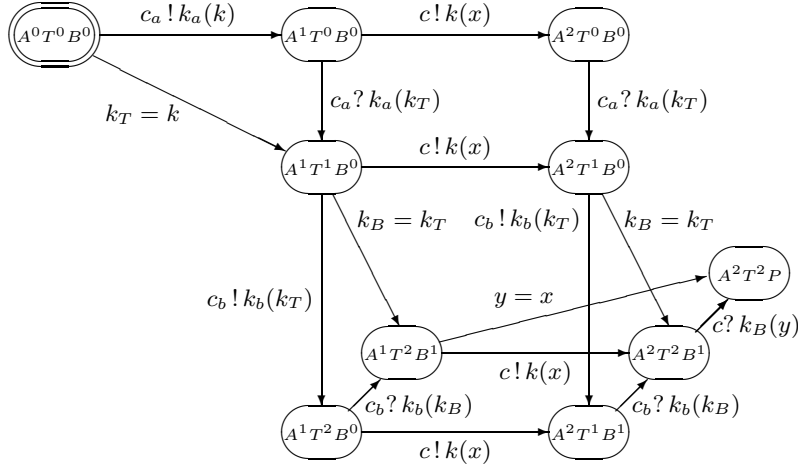
$$\by
\begin{picture}(270,160)

\put(0,150){\oval(30,20)}
\put(0,150){\oval(34,24)}
\put(0,150){\makebox(0,0)[c]{${\scriptstyle A^0T^0B^0}$}}

\put(100,150){\oval(30,20)}
\put(100,150){\makebox(0,0)[c]{${\scriptstyle A^1T^0B^0}$}}

\put(100,100){\oval(30,20)}
\put(100,100){\makebox(0,0)[c]{${\scriptstyle A^1T^1B^0}$}}

\put(100,0){\oval(30,20)}
\put(100,0){\makebox(0,0)[c]{${\scriptstyle A^1T^2B^0}$}}

\put(200,150){\oval(30,20)}
\put(200,150){\makebox(0,0)[c]{${\scriptstyle A^2T^0B^0}$}}

\put(200,100){\oval(30,20)}
\put(200,100){\makebox(0,0)[c]{${\scriptstyle A^2T^1B^0}$}}

\put(200,0){\oval(30,20)}
\put(200,0){\makebox(0,0)[c]{${\scriptstyle A^2T^1B^1}$}}

\put(100,125){\makebox(0,0)[l]{
$c_a?\, k_a(k_T)$
}}

\put(200,125){\makebox(0,0)[l]{
$c_a?\, k_a(k_T)$
}}

\put(100,50){\makebox(0,0)[r]{
$c_b\,!\, k_b(k_T)$
}}

\put(200,80){\makebox(0,0)[r]{
$c_b\,!\, k_b(k_T)$
}}

\put(50,153){\makebox(0,0)[b]{
$c_a\,!\, k_a( k)$
}}

\put(150,153){\makebox(0,0)[b]{
$c\,!\, k(x)$
}}

\put(150,103){\makebox(0,0)[b]{
$c\,!\, k(x)$
}}

\put(150,-3){\makebox(0,0)[t]{
$c\,!\, k(x)$
}}

\put(50,120){\makebox(0,0)[r]{
$ k_T=k$
}}

\put(100,140){\vector(0,-1){30}}
\put(200,140){\vector(0,-1){30}}

\put(100,90){\vector(0,-1){80}}
\put(200,90){\vector(0,-1){80}}

\put(14.5,142){\vector(2,-1){72}}

\put(17,150){\vector(1,0){68}}

\put(115,150){\vector(1,0){70}}
\put(115,100){\vector(1,0){70}}
\put(115,0){\vector(1,0){70}}

\put(130,30){\oval(30,20)}
\put(130,30){\makebox(0,0)[c]{${\scriptstyle A^1T^2B^1}$}}


\put(230,30){\oval(30,20)}
\put(230,30){\makebox(0,0)[c]{${\scriptstyle A^2T^2B^1}$}}

\put(180,27){\makebox(0,0)[t]{
$c\,!\, k(x)$
}}


\put(145,30){\vector(1,0){70}}


\put(260,60){\oval(30,20)}
\put(260,60){\makebox(0,0)[c]{${\scriptstyle A^2T^2P}$}}

\put(110,8){\vector(1,1){12}}

\put(210,8){\vector(1,1){12}}
\put(240,38){\vector(1,1){12}}
\put(240,38){\vector(1,1){12}}

\put(205,90){\vector(1,-2){25}}
\put(104,90){\vector(1,-2){25}}

\put(144,34){\vector(4,1){100}}

\put(113,10){\makebox(0,0)[l]{
$c_b?\, k_b(k_B)$
}}


\put(213,10){\makebox(0,0)[l]{
$c_b?\, k_b(k_B)$
}}
\put(243,40){\makebox(0,0)[l]{
$c?\,k_B(y)$
}}

\put(110,80){\makebox(0,0)[l]{
$k_B=k_T$
}}
\put(210,80){\makebox(0,0)[l]{
$k_B=k_T$
}}

\put(190,50){\makebox(0,0)[r]{
$y=x$
}}
\end{picture}
\ey
$$
 }
}

It is not so difficult to prove that
the property 
$b_{A^2T^2P}\leq (x=y)$
and the equivalence
$[y=x]\,[y=x]\,P\approx
[y=x]\,P$
imply 
$Sys\approx \tilde Sys$.

The secrecy property is a 
direct consequence of the 
integrity property.
$\blackbox$

\end{document}